\documentclass[10pt]{article}
\usepackage{times}

\usepackage{fullpage}
\usepackage[T1]{fontenc}
\usepackage{amsmath}
\usepackage{amssymb}
\usepackage{latexsym}
\usepackage{bussproofs}

\usepackage{array}
\usepackage{alltt}
\usepackage{theorem}

\title{A finiteness structure on resource terms}
\author{Thomas Ehrhard\thanks{This work has been
partly funded by the ANR project BLAN07-1 189926 \emph{Curry-Howard
for Concurrency} (CHOCO).}\\
Preuves, Programmes, Systèmes\\
CNRS and Université Paris Diderot - Paris 7\\
\texttt{Thomas.Ehrhard@pps.jussieu.fr}
}

\newtheorem{theorem}{Theorem}
\newtheorem{proposition}[theorem]{Proposition}
\newtheorem{lemma}[theorem]{Lemma}

\theorembodyfont{\normalfont}
\theoremstyle{plain}

\def\bull{\vrule height.9ex width.8ex depth-.1ex}

{\smallbreak\noindent{\bf Proof: }\nobreak}%
{\hbox{}\hfill\bull\normalsize\smallbreak}

\newcommand{\proofitem}[1]{\paragraph{\mdseries\textit{#1}}}

\newcommand{\Beginproof}{\proofitem{Proof.}}
\newcommand{\Endproof}{
  \ifmmode 
  \else \leavevmode\unskip\penalty9999 \hbox{}\nobreak\hfill
  \fi
  \quad\hbox{$\Box$}
  \par\medskip}

\newenvironment{remark}%
{\smallbreak\noindent{\textit{Remark\/}: }\nobreak}%
{\smallbreak}

\renewcommand{\phi}{\varphi}
\renewcommand\epsilon{\varepsilon}

\newcommand{\Implies}{\Rightarrow}

\newcommand{\St}{\mid}

\newcommand{\Bot}{{\mathord{\perp}}}

\newcommand\cF{\mathcal{F}}

\newcommand\cI{\mathcal{I}}

\newcommand\cP{\mathcal{P}}

\newcommand\cS{\mathcal{S}}

\newcommand\cV{\mathcal{V}}

\newcommand\Fini{{\mathrm{fin}}}

\newcommand\Part[1]{{\cal P}({#1})}

\newcommand\Myleft{}
\newcommand\Myright{}

\newcommand\Web[1]{\Myleft|{#1}\Myright|}
\newcommand\Supp[1]{\operatorname{\mathsf{Supp}}({#1})}

\newcommand\Orth[1]{{#1}^\Bot}

\newcommand\Biorth[1]{{#1}^{\Bot\Bot}}

\newcommand\Seq[1]{\vdash{#1}}

\newcommand\Partfin[1]{{\cP_\Fini}({#1})}

\newcommand\Locun[1]{1^J}

\newcommand\Isom\simeq

\newcommand\Nat{{\mathbb{N}}}

\newcommand\Biind[2]{\genfrac{}{}{0pt}{1}{#1}{#2}}

\newcommand\Fin[1]{\mathsf{F}({#1})}

\newcommand\Prom[1]{{#1}^!}

\newcommand\App[2]{\left({#1}\right){#2}}
\newcommand\Abs[2]{\lambda{#1}\,{#2}}

\newcommand\List[3]{#1_{#2},\dots,#1_{#3}}

\newcommand\Subst[3]{{#1}\left[{#2}/{#3}\right]}

\newcommand\Factor[1]{{#1}!}

\newcommand\Ring{R}
\newcommand\Linapp[2]{\langle{#1}\rangle{#2}}
\newcommand\Fmod[2]{{#1}\langle{#2}\rangle}
\newcommand\Fmodfin[2]{{#1}\langle{#2}\rangle}
\newcommand\Fmodrel[2]{{#1}\langle{#2}\rangle}

\newcommand\Imod[2]{{#1}\langle{#2}\rangle_\infty}
\newcommand\FImod[2]{{#1}\langle{#2}\rangle_{(\infty)}}

\newcommand\Mfin[1]{\mathcal M_\Fini({#1})}

\newcommand\Neigh[2]{\operatorname{V}_{#1}(#2)}

\newcommand\Sterms{\Delta}
\newcommand\Pterms{\Delta^{!}}

\newcommand\Nsterms{\Delta_0}

\newcommand\Rel[1]{\mathrel{#1}}

\newcommand\Redone{\beta^1_\Sterms}
\newcommand\Redonep{\beta^1_{\Pterms}}
\newcommand\Redonez{\beta^{0,1}_\Sterms}
\newcommand\Redonezp{\beta^{0,1}_{\Pterms}
\newcommand\Redmult{\beta^{\mathrm{m}}_\Delta}}
\newcommand\Red{\beta_\Sterms}
\newcommand\Redp{\beta_{\Pterms}}
\newcommand\Redst[1]{\mathop{\mathsf{Red}}}

\newcommand\Shape[1]{\mathcal{T}(#1)}
\newcommand\Tay[1]{{#1}^*}

\newcommand\Redeq{=_\Delta}

\newcommand\NF{\operatorname{\mathsf{NF}}}

\newcommand\Msubst[3]{\partial_{#3}(#1,#2)}
\newcommand\Symgrp[1]{\mathfrak S_{#1}}

\newcommand\Size[1]{\mathsf S({#1})}

\newcommand\Msetofsubst[1]{\bar F}

\newcommand\FV{\operatorname{\mathrm{FV}}}

\newcommand\Down[1]{\mathord\downarrow{#1}}
\newcommand\Up[1]{\mathord\uparrow{#1}}
\newcommand\Snfinite{\operatorname{\mathcal{N}_1}}
\newcommand\Snfinitefv{\operatorname{\mathcal{N}}}
\newcommand\Snfinitevar{\operatorname{\mathcal{N}}_0}
\newcommand\Pnfinite{\Snfinite^!}

\newcommand\Field{\mathbf{k}}
\newcommand\Snneigh[1]{\operatorname{\mathsf V}({#1})}

\newcommand\Impl[2]{{#1}\Rightarrow{#2}}
\newcommand\Preimpl[2]{{#1}\bullet{#2}}

\newcommand\Tforall[2]{\forall #1\,#2}
\newcommand\Alambda{\Lambda_\Field}

\newcommand\Tsem[2]{[#1]^{#2}}
\newcommand\Satfs{\textsf{SFS}}
\newcommand\Setsr{\mathbb S}

\begin{document}
\maketitle
\begin{abstract}
  In our paper "Uniformity and the Taylor expansion of ordinary lambda-terms"
  (with Laurent Regnier), we studied a translation of lambda-terms as infinite
  linear combinations of resource lambda-terms, from a calculus similar to
  Boudol's lambda-calculus with resources and based on ideas coming from
  differential linear logic and differential lambda-calculus. The good
  properties of this translation wrt.\ beta-reduction were guaranteed by a
  coherence relation on resource terms: normalization is "linear and stable"
  (in the sense of the coherence space semantics of linear logic) wrt.\ this
  coherence relation. Such coherence properties are lost when one considers
  non-deterministic or algebraic extensions of the lambda-calculus (the
  algebraic lambda-calculus is an extension of the lambda-calculus where terms
  can be linearly combined). We introduce a "finiteness structure" on resource
  terms which induces a linearly topologized vector space structure on terms
  and prevents the appearance of infinite coefficients during reduction, in
  typed settings.
\end{abstract}

\section*{Introduction}

\paragraph{Denotational semantics and linear logic.}
Denotational semantics consists in interpreting syntactical objects (programs,
proofs) as points in abstract structures (typically, ordered sets with various
completeness properties). In this process, the dynamical features of programs
are lost, and abstract properties of programs, such as continuity, stability or
sequentiality are expressed. 

A program, or a proof, is normally a finite object, and its denotation is
usually infinite, because it describes all the possible behaviors of the
program when applied to all possible arguments. Semantics turns the potential
infinity of program dynamics into the actually infinite static description of
all its potential behaviors.

Linear logic (LL), which arose from investigations in denotational semantics,
sheds a new light on this picture. Whilst being as expressive as intuitionistic
logic, LL contains a purely linear fragment which is completely finite in
the sense that, during reduction, the size of proofs strictly decreases. For
allowing to define and manipulate potentially infinite pieces of
proofs/programs, LL introduces new connectives: the exponentials.

Unlike its finite multiplicative-additive fragment, the exponential fragment of
LL is strongly asymmetric:
\begin{itemize}
\item on one side, there is a \emph{promotion} rule which allows to introduce
  the ``$!$'' connective and makes a proof duplicable and erasable;
\item and on the other side, there are the rules of \emph{contraction},
  \emph{weakening} and \emph{dereliction} which allow to duplicate, erase
  and access to promoted proofs. These rules introduce and allow to perform
  deductions on the ``$?$'' connective, which is the linear dual of
  ``$!$''. Let use call these rules \emph{structural}\footnote{It is not really
    standard to consider dereliction as structural.}.
\end{itemize}
The only infinite rule of LL is promotion. The potentially infinite duplicating
power of contraction is not ``located'' in the contraction rule itself, but in
the fact that, for being duplicable by contractions, a proof must be promoted
first. This fact can be observed in denotational models but is not clear in the
syntax because the structural rules have no other opponents but
promotion\footnote{This picture is not completely faithful because promotion
  has also to be considered as a ``$?$'' rule.}.

\paragraph{Differential linear logic}
The situation is quite different in differential LL (and, implicitly, in
differential lambda-calculus and its variants), a system that we introduced
recently (see~\cite{EhrhardRegnier02,EhrhardRegnier06d,EhrhardLaurent08}). In
this system, the ``$?$'' rules have exact dual rules: there is a
\emph{cocontraction}, a \emph{coweakening} and a \emph{codereliction}
rules. These rules are logical versions of standard mathematical operations
used in elementary differential calculus, whence the name of the system.

So in differential LL we have structural and costructural rules and these rules
interact in a completely symmetric and \emph{finite} way, just as in the
multiplicative and additive fragment. Promotion remains apart, as the only
truly infinite rule of logic. This fact, which in LL could be observed only in
denotational models, can be expressed syntactically in differential LL by
means of the Taylor expansion of promotion rules.

\paragraph{Resource lambda-calculus.}
This operation is more easily understood in the lambda-calculus
(see~\cite{Tranquilli08} for the connection between lambda-terms and nets in
differential LL). Roughly speaking, the ordinary lambda-calculus correspond to
the fragment of LL which contains the multiplicative, structural and promotion
rules. But we can also consider a lambda-calculus corresponding to the
multiplicative, structural and costructural rules: the resource calculus that
we introduced in~\cite{EhrhardRegnier06a}. Similar calculi already existed in
the literature, such as Boudol's calculi with multiplicities~\cite{Boudol93} or
with resources~\cite{BoudolCurienLavatelli99}, and also Kfoury's
calculi~\cite{Kfoury00}, introduced with different motivations and with
different semantic backgrounds. The intuition behind our calculus with
resources is as follows.

The first thing to say is that types should be thought of as (topological)
vector spaces and not as domains. Consider then a term $t:\Impl AB$ which
should be seen as a function from $A$ to $B$. Then imagine that it makes sense
to compute the $n$-the derivative of $t$ at the point $0$ of the vector space
$A$: it is a function $t^{(n)}(0):A^n\to A$, separately linear in each of its
argument, and symmetric in the sense that
$t^{(n)}(0)(s_1,\dots,s_n)=t^{(n)}(0)(s_{f(1)},\dots,s_{f(n)})$ for any
permutation $f\in\Symgrp n$ and any tuple $(s_1,\dots,s_n)\in A^n$. In our
resource calculus, we have an application construction which represents this
operation. Given a term $t$ (of type $\Impl AB$ if we are in a typed setting)
and a finite number $\List s1n$ of terms (of type $A$), we can ``apply'' $t$ to
the multiset $S=s_1\cdots s_n$ (the multiset whose elements are $\List s1n$,
taking multiplicities into account) and we denote with $\Linapp sS$ this
operation. We take benefit of the intrinsic commutativity of multisets for
implementing the symmetry of the $n$-th derivative. The other constructions of
this calculus are standard: we have variables $x,y,\dots$ and abstractions
$\Abs xs$. Redexes are terms of the shape $\Linapp{\Abs xs}S$ and $x$ can have
several free occurrences in $s$, which \emph{are all linear}. When reducing
this redex, one does not duplicate $S$. Instead, one splits it into as many
pieces as there are occurrences of $x$ in $s$, and since all these occurrences
are linear, all these pieces should contain exactly one term. We do that in all
possible ways and take the sum of all possible results. When the number of free
occurrences of $x$ in $s$ and the size of $S$ do not coincide, the result of
this operation is $0$.

For this to make sense, one must have the possibility of adding terms, and this
is compatible with the idea that types are vector spaces.

\paragraph{Taylor expansion.}
Taylor expansion consists in replacing the ordinary application of
lambda-calculus with this differential application of the resource calculus. If
$M:\Impl AB$ and $N:A$ are terms, then the standard Taylor formula should be
\begin{equation*}
  \App MN=\sum_{n=0}^\infty
  \frac 1{\Factor n} M^{(n)}(0)(\overbrace{N,\dots,N}^n)
\end{equation*}
This leads to the idea of writing any term $M$ as an infinite linear
combination of resource terms (with rational coefficients): if $\Tay M$ and
$\Tay N$ are such sums, we should have
\begin{equation}\label{eq:taylor-product}
  \Tay{(\App MN)}=\sum_{n=0}^\infty
  \frac 1{\Factor n} \Linapp{\Tay M}{(\Tay N)^n}
\end{equation}
where the power $(\Tay N)^n$ has to be understood in the sense of multiset
concatenation, extented to linear combinations of multisets by linearity. Using
the fact that all the constructions of the resource calculus should be linear
(that is, should distribute over arbitrary linear combinations),
formula~(\ref{eq:taylor-product}) leads to a definition of $\Tay M$ as a linear
combination of resource terms: $\Tay M=\sum_{s\in\Sterms}\Tay M_s s$ where each
$\Tay M_s$ is a positive rational number ($\Sterms$ is the set of resource
terms): this is the Taylor expansion of $M$.

Taylor expansion looks like denotational semantics: we have transformed a
finite program $M$ with a rich, potentially infinite, dynamics into an infinite
set (linear combination to be more precise) of more elementary things, the
resource terms. The difference wrt.\ denotational semantics is that these terms
have still a dynamics, but this dynamics is completely finite because they
belong to the promotion-free fragment of differential linear logic: all terms
of our resource calculus, even the non typeable ones, are trivially strongly
normalizing. But of course there is no uniform bound on the length of the
reductions of the resource terms appearing in the Taylor expansion of a term.

\paragraph{Content.}
The present article is a contribution to a programme which consists in
considering infinite linear combinations of resource terms as generalized
lambda-terms. The first point to understand is how beta-reduction can be
applied to such infinite linear combinations without introducing infinite
coefficients.  We initiated this programme in~\cite{EhrhardRegnier06a},
defining a binary symmetric, but not reflexive, coherence relation on resource
terms (such a coherence relation has also been defined for differential
interaction nets in~\cite{PaganiTasson09}) and showing that, if two terms $s$
and $t$ are coherent and distinct, then their normal forms are disjoint (and
hence can be summed). So a first idea is to consider cliques as generalized
lambda-terms, and this is sound because the resource terms appearing in the
Taylor expansion of a lambda-term are pairwise coherent.

But if we allow linear combinations in the lambda-calculus (as in the
differential lambda-calculus for instance, and we speak then of algebraic
lambda-calculus\footnote{There are other algebraizations of the
  lambda-calculus, we think in particular of the calculus considered by Arrighi
  and Dowek~\cite{ArrighiDowek08} which is quite different from ours because
  application is right-linear in their setting}), then we cannot expect Taylor
expansions to be cliques for that coherence relation. Instead, we equip the set
of resource terms with a finiteness structure (in the sense
of~\cite{Ehrhard00b}) which is defined in such a way that for any ``finitary''
linear combination $\sum_s\alpha_s s$ of resource lambda-terms, the sum
$\sum_s\alpha_sa_s$ always makes sense, whatever be the choices of $a_s$ such
that $s$ beta-reduces to $a_s$ in the resource lambda-calculus. We prove a
soundness theorem, showing that the Taylor expansion of an algebraic
lambda-terms is always finitary. This cannot hold however for the untyped
algebraic lambda-calculus because we know that this calculus leads to unbounded
coefficients during beta-reduction (think of $\App\Theta{\Abs x{(z+x)}}$ where
$z\not=x$ and $\Theta$ is the Turing fixpoint combinator). So we prove our
soundness result for second-order typeable algebraic lambda-terms, by a method
similar to Girard's proof of strong normalization of system F in Krivine's very
elegant presentation~\cite{Krivine93}. The method consists in associating with
any type a finiteness space (and hence a linearly topologized vector space)
whose underlying set (web) is a set of resource terms.

\section{The resource lambda-calculus}

\subsection{The calculus}

The syntax of our resource calculus is defined as follows. One defines first
the set $\Sterms$ of simple terms and the set $\Pterms$ of simple poly-terms.
\begin{itemize}
\item If $x$ is a variable then $x\in\Sterms$;
\item if $s\in\Sterms$ and $x$ is a variable then $\Abs xs\in\Sterms$;
\item if $s\in\Sterms$ and $S\in\Pterms$ then $\Linapp sS\in\Sterms$;
\item if $s_1,\dots,s_n\in\Sterms$ then the multiset which consists of the
  $s_i$s, denoted in a multiplicative way as $s_1\cdots s_n$, is an
  element of $\Pterms$. The empty simple poly-term is accordingly denoted as
  $1$.
\end{itemize}

We define the size $\Size s$ of a simple term $s$ and the size $\Size S$ of a
simple poly-term by induction as follows:
\begin{itemize}
\item $\Size x=1$
\item $\Size{\Abs xs}=1+\Size s$
\item $\Size{\Linapp sS}=1+\Size s+\Size S$
\item $\Size{s_1\cdots s_2}=\Size{s_1}+\cdots+\Size{s_n}$.
\end{itemize}

\subsubsection{Extended syntax.}\label{par:extended-syntax}

Given a rig (semi-ring) $R$ and a set $E$, we denote by $\Fmod RE$ the set of
all formal finite linear combinations of elements of $E$ with coefficients in
$R$: it is the free $R$-module generated by $E$. If $a\in\Fmod RE$ and $s\in
E$, $a_s\in R$ denotes the coefficient of $s$ in $a$. We also define $\Imod RE$
as the set of all (not necessarily finite) linear combinations of elements of
$E$ with coefficients in $R$; we use the same notations as for the elements of
$\Fmod RE$ and we use $\FImod\Ring E$ to denote both modules, to deal with
constructions which are applicable in both settings.

The semi-rings that we consider are
\begin{itemize}
\item $\Setsr=\{0,1\}$ with $1+1=1$, so that $\Fmod\Setsr E=\Partfin E$ and
  $\Imod\Setsr E=\Part E$;
\item $\Nat$, and then $\Fmod\Nat E$ is the set of all finite multisets of
  elements of $E$. Given $a\in \Fmod\Nat E$ and $s\in E$ we write $s\in a$ when
  $a_s\not=0$;
\item a field $\Field$, and then $\Fmod\Field E$ is the $\Field$-vector space
  generated by $E$ and $\Imod\Field E$ is also a vector space.
\end{itemize}

Let $a\in\FImod R\Sterms$, we set \(\Abs xa=\sum_{s\in\Sterms}a_s\Abs
xs\in\FImod R\Sterms\).  Given moreover $A\in\FImod R\Pterms$, we set \(\Linapp
aA=\sum_{s\in\Sterms, S\in\Pterms}a_sA_S\Linapp sS\in\FImod R\Sterms\). Last,
given $a(1),\dots,a(n)\in\FImod R\Sterms$, we define $a(1)\cdots a(n)$ as
\(\sum_{s(1),\dots,s(n)\in\Sterms}a(1)_{s(1)}\cdots a(n)_{s(n)}(s(1)\cdots
s(n))\in\FImod R\Pterms\). In that formula, remember that $s(1)\cdots s(n)$ is
the multiset made of $s(1),\dots,s(n)$. This formula expresses that we consider
multiset concatenation as a product, and so, when extended to linear
combinations, a distributivity law must hold.

In particular, given $a\in\FImod R\Sterms$ and $n\in\Nat$, we set
$a^n=\overbrace{a\cdots a}^n\in\FImod R\Pterms$. When $R=\Field$, we set $\Prom
a=\sum_{n\in\Nat}\frac 1{\Factor n}a^n\in\Imod\Field\Pterms$ (this sum always
makes sense, and we require $R=\Field$ to give a meaning to $1/\Factor n$). For
$e\subseteq\Sterms$ (that is $e\in\Imod\Setsr\Sterms$), we set $\Prom e=\Mfin
e\subseteq\Pterms$.

So all the constructions of the syntax can be applied to arbitrary linear
combinations of simple terms, giving rise to combinations of simple terms.

\subsubsection{Differential substitution}\label{par:diff-subst}

Given $s\in\Sterms$ and $S\in\Pterms$, and given a variable $x$, we define the
differential substitution $\Msubst sSx$ as $0$ if the number of free
occurrences of $x$ in $s$ is different from $n$, and as $\sum_{f\in\Symgrp
  n}s[s_{f(1)}/x_1,\dots,s_{f(n)}/x_n]$ otherwise, where $S=s_1\cdots s_n$,
$x_1,\dots,x_n$ are the $n$ occurrences of $x$ in $s$ and $\Symgrp n$ is the
group of permutations on $\{1,\dots,n\}$.

Given $s\in\Sterms$ and $S_1,\dots,S_n\in\Pterms$ and pairwise distinct
variables $\List x1n$ which do not occur free in the $S_i$'s, we define more
generally the parallel differential substitution $\Msubst{s}{\List S1n}{\List
  x1n}$: the definition is similar (the sum is indexed by tuples
$(f_1,\dots,f_n)$ where $f_i$ is a permutation on the free occurrences of $x_i$
in $s$).

This operation must be extended by linearity. Given $a\in\FImod R\Sterms$ and
$A\in\FImod R\Pterms$, we set
\begin{equation*}
  \Msubst{a}{A}{x}
  =\sum_{s\in\Sterms,S\in\Pterms}a_sA_S\Msubst sSx\in\FImod R\Sterms
\end{equation*}
and we define similarly $\Msubst{a}{\List A1n}{\List x1n}\in\FImod
R\Sterms$. It is not obvious at first sight that this sum is well defined in
the infinite case. This results from Lemma~\ref{lemma:msubst-reverse} (see
below).

\subsubsection{The reduction relations}

Given two sets $E$ and $F$ and a relation $\rho\subseteq E\times\Fmod\Nat F$,
we define a relation $\Fmodrel\Nat\rho\subseteq\Fmod\Nat E\times\Fmod\Nat F$ as
follows: we say that $(a,b)\in\Fmodrel\Nat\rho$ if there are
$(s_1,a_1),\dots,(s_n,a_n)\in\rho$ such that $s_1+\cdots+s_n=a$ and
$b_1+\cdots+b_n=b$.

The one step reduction relations
$\Redone\subseteq\Sterms\times\Fmod\Nat\Sterms$ and
$\Redonep\subseteq\Pterms\times\Fmod\Nat\Pterms$ are defined as follows.
\begin{itemize}
\item $x\Rel\Redone b$ never holds;
\item $\Abs xs\Rel\Redone b$ if $b=\Abs xa$ with $s\Rel\Redone a$;
\item $s_1\cdots s_n\Rel\Redonep B$ if, for some $i$, $s_i\Rel\Redone b_i$ and
  $B=s_1\cdots b_i\cdots s_n$;
\item $\Linapp sS\Rel\Redone b$ in one of the following situations
  \begin{itemize}
  \item $s\Rel\Redone a$ and $b=\Linapp aS$;
  \item $S\Rel\Redonep A$ and $b=\Linapp sA$;
  \item $s=\Abs xt$ and $b=\Msubst tSx$.
  \end{itemize}
\end{itemize}

\begin{lemma}\label{lemma:redone-decroit}
  Let $s\in\Sterms$ and $b\in\Fmod\Nat\Sterms$. If $s\Rel\Redone b$, then, for
  any $t_1,t_2\in b$, one has $\Size{t_1}=\Size{t_2}<\Size s$.
\end{lemma}
The proof is straightforward (simple case inspection).

Let $\Redonez=\Fmodrel\Nat{\{(s,s)\St s\in\Sterms\}\cup\Redone}$ and
$\Redonezp=\Fmodrel\Nat{\{(S,S)\St s\in\Pterms\}\cup\Redonep}$. These are
reflexive reduction relations on $\Fmod\Nat\Sterms$ and $\Fmod\Nat\Pterms$
respectively. More explicitly, we have $a\Rel\Redonez b$ if one can write
$a=s_1+\cdots+s_n+a'$ and $b=b_1+\cdots+b_n+a'$ with $s_i\Rel\Redone b_i$ for
$i=1,\dots,n$, and similarly for $\Redonezp$.

Finally we denote with $\Red$ and $\Redp$ respectively the transitive closures
of these relations.

\begin{lemma}\label{lemma:redone-subst}
  Let $s,t\in\Sterms$ and $x$ is a variable which occurs free exactly once in
  $s$. If $s\Rel\Redone a$ then $\Subst stx\Rel{\Redone}\Subst atx$ and if
  $t\Rel\Redone b$ then $\Subst stx\Rel{\Redone}\Subst sbx$.
\end{lemma}

\begin{lemma}\label{lemma:redone-msubst}
  If $s\Rel\Redone a$ then $\Msubst sSx\Rel{\Fmodrel\Nat\Redone}\Msubst
  aSx$. If $S\Rel\Redone A$ then $\Msubst sSx\Rel{\Fmodrel\Nat\Redone}\Msubst
  sAx$. 
\end{lemma}
These two lemmas are proved by straightforward inductions.

The reduction relation $\Red$ on $\Fmod\Nat\Sterms$ has good properties: it is
strongly normalizing, confluent
(see~\cite{EhrhardRegnier02,Vaux05,PaganiTranquilli09}). Given $s\in\Sterms$,
we denote by $\NF(s)$ the unique normal form of $s$, which is an element of
$\Fmod\Nat\Sterms$.

\subsubsection{Examples of reduction}
Of course $\Linapp{\Abs xx}{y}\Rel\Red y$, but if the identity is applied to a
multiset of size $\not=1$, the result is $0$: $\Linapp{\Abs xx}1\Rel\Red 0$ and
$\Linapp{\Abs xx}{y^2}\Rel\Red 0$ (where $y^2$ is the multiset which contains
twice the variable $y$; this notation is compatible with the distributivity
laws of~\ref{par:extended-syntax}).

Similarly, the term $\Linapp x{x^2}$ contains $3$ occurrences of $x$ (it is
sensible to say that it is of degree $3$ in $x$). So $\Linapp{\Abs x{\Linapp
    x{x^2}}}S\Rel\Red 0$ if the size of $S$ is $\not=3$. And we have
$\Linapp{\Abs x{\Linapp x{x^2}}}{(y^2z)}\Rel\Red\Msubst{\Linapp
  x{x^2}}{y^2z}{x}=4\Linapp y{yz}+2\Linapp z{y^2}$. As a last example we have
$\Linapp{\Abs x{\Linapp{\Linapp x{x}}x}}{(y^2z)}\Rel\Red\Msubst{\Linapp
  x{x^2}}{y^2z}{x}=2\Linapp{\Linapp yz}y+2\Linapp{\Linapp yy}z+2\Linapp{\Linapp
  zy}y$.

\subsubsection{An order relation on simple terms and poly-terms.}
Let us define an order relation on simple terms. Given
$s,t\in\Sterms$, we write $t\leq s$ if there exists $a\in\Fmod\Nat\Sterms$
such that $s\Rel\Red a$ and $t\in a$. Given $s\in\Sterms$, we use
$\Down s=\{t\in\Sterms\St t\leq s\}$ and $\Up s=\{t\in\Sterms\St t\geq s\}$. We
define similarly an order relation on poly-terms and introduce similar
notations: $T\leq S$, $\Up S$ and $\Down S$.

\begin{lemma}\label{lemma:down-finite}
  For any $s\in\Sterms$, the set $\Down s$ is finite.  
\end{lemma}
\Beginproof
By Lemma~\ref{lemma:redone-decroit} and König's lemma.
\Endproof

\subsection{Two technical lemmas}

\begin{lemma}\label{lemma:red-permutation}
  Let $y$ be a variable and $S_1,\dots,S_n\in\Pterms$ which do not contain free
  the variable $y$ and let $v=\Linapp{\cdots\Linapp{\Linapp
      y{S_1}}{S_2}\cdots}{S_n}$. Let $s\in\Sterms$, $S\in\Pterms$, $x$ be a
  variable. Let $t\in\Sterms$ be such that $t\leq\Subst{v}{\Linapp{\Abs
      xs}S}y$. Then one of the the two following cases arises:
  \begin{itemize}
  \item either $t=\Subst{v'}{\Linapp{\Abs x{s'}}{S'}}y$ with $v'\leq v$,
    $s'\leq s$ and $S'\leq S$
  \item or $t\leq\Subst vuy$ for some $u\in\Msubst sSx$.
  \end{itemize}
\end{lemma}
\Beginproof
By induction on $\Size v+\Size s+\Size S$.  Let $b\in\Fmod\Nat\Sterms$ be such
that $\Subst v{\Linapp{\Abs xs}{S}}y\Rel\Red b$ and $t\in b$. Consider the
first reduction step of this reduction. Four cases are possible, because of the
particular shape of $v$.

First case: the reduction occurs in $s$. That is $s\Rel\Redone a$ for some
$a\in\Fmod\Nat\Sterms$ and the reduction $\Subst v{\Linapp{\Abs
    xs}{S}}y\Rel\Red b$ splits in $\Subst v{\Linapp{\Abs
    xs}{S}}y\Rel\Redone\Subst v{\Linapp{\Abs xa}{S}}y\Rel\Red b$. Since $t\in
b$, one can find some $u\in\Sterms$ with $u\in a$ such that $t\leq\Subst
v{\Linapp{\Abs x{u}}{S}}y$. Since $\Size{u}<\Size s$, the inductive hypothesis
applies and so there are two cases.
\begin{itemize}
\item Either we have $t=\Subst{v'}{\Linapp{\Abs x{u'}}{S'}}y$ with $v'\leq v$,
  $u'\leq u$ and $S'\leq S$ and we conclude because $u<s$.
\item Or $t\leq\Subst{v}{w}y$ with $w\in\Sterms$ such that $w\in\Msubst uSx$.
  Since $u\in a$ and $w\in\Msubst uSx$, we have $w\in\Msubst aSx$.  But
  $\Msubst sSx\Rel{\Fmodrel\Nat\Redone}\Msubst aSx$ by
  Lemma~\ref{lemma:redone-msubst} and hence there exists $w_0\in\Msubst sSx$
  such that $w<w_0$. Hence we have $\Subst vwy<\Subst v{w_0}y$ by
  Lemma~\ref{lemma:redone-subst} and we conclude by transitivity.
\end{itemize}

The second case, where the reduction occurs in $S$ is similar.

Third case: the reduction occurs in $v$. That is $v\Rel\Redone
c\in\Fmod\Nat\Sterms$ and the reduction $\Subst v{\Linapp{\Abs xs}{S}}y\Rel\Red
b$ splits in $\Subst v{\Linapp{\Abs xs}{S}}y\Rel\Redone\Subst c{\Linapp{\Abs
    xs}S}y\Rel\Red b$. Since $t\in b$, one can find some $w\in c$ such that
$t\leq\Subst w{\Linapp{\Abs xs}{S}}y$. Since $\Size w<\Size v$, the inductive
hypothesis applies and so there are two cases.
\begin{itemize}
\item Either $t=\Subst{w'}{\Linapp{\Abs x{s'}}{S'}}y$ with $w'\leq w$, $s'\leq
  s$ and $S'\leq S$ and we conclude because $w\leq v$.
\item Or $t\leq\Subst wuy$ for some $u\in\Msubst sxS$. We conclude by
  Lemma~\ref{lemma:redone-subst} because $w<v$.
\end{itemize}

Last case: the reduction $\Subst v{\Linapp{\Abs xs}{S}}y\Rel\Red b$ splits in
$\Subst v{\Linapp{\Abs xs}{S}}y\Rel\Redone\Subst v{\Msubst sSx}y\Rel\Red b$ and
we conclude immediately that there exists $u\in\Msubst sSx$ such that
$t\leq\Subst vux$.
\Endproof

\begin{lemma}\label{lemma:msubst-reverse}
  Let $s\in\Sterms$. There are only finitely many pairs
  $(t,T)\in\Sterms\times\Pterms$ such that $s\in\Msubst tTx$.
\end{lemma}
\Beginproof
(Sketch) The intuition is clear and can easily be formalized. For building
$(t,T)$, one must choose some $n\in\Nat$, and then $n$ \emph{pairwise
  disjoint}\footnote{None of these terms can be a sub-term of another one.}
sub-terms $\List t1n$ of $s$. Then $t$ is obtained by replacing these sub-terms
by $x$ in $s$, and $T=t_1\cdots t_n$. There are only finitely many ways of
choosing such a tuple $(n,\List t1n)$.
\Endproof

\section{Finiteness spaces}\label{sec:finiteness-spaces}
We recall some basic material on finiteness spaces. Given a set $I$ and a
collection $\cF$ of subsets of $I$, we define
\begin{equation*}
  \Orth\cF=\{e'\subseteq I\St\forall e\in\cF\ e\cap e'\ \text{is finite}\}\,.
\end{equation*}
A finiteness space is a pair $X=(\Web X,\Fin X)$ where $\Web X$ is a set (the
web of $X$) and $\Fin X\subseteq\Part{\Web X}$ satisfies $\Biorth{\Fin
  X}\subseteq\Fin X$ (the other inclusion being always true). The following
properties follow immediately from this definition: if $e\subseteq\Web X$ is
finite then $e\in\Fin X$; if $e\in\Fin X$ and $f\subseteq e$ then $f\in\Fin X$;
if $e_1,e_2\in\Fin X$ then $e_1\cup e_2\in\Fin X$.

\paragraph{Vector space.}
Let $\Field$ be a field.  Given $a\in\Field^{\Web X}$, let $\Supp a=\{s\in\Web
X\St a_s\not=0\}$ (the support of $a$). We set $\Fmod\Field
X=\{a\in\Field^{\Web X}\St\Supp a\in\Fin X\}$. This set is a $\Field$-vector
space, addition and scalar multiplication being defined pointwise.

\paragraph{Topology.}
Given $e'\in\Orth{\Fin X}$, let $\Neigh{0}{e'}=\{a\in\Fmod\Field X\St\Supp
a\cap e'=\emptyset\}$: this is a linear subspace of $\Fmod\Field X$. A subset
$\cV$ of $\Fmod\Field X$ is open if, for all $a\in\cV$ there exists
$e'\in\Orth{\Fin X}$ such that $a+\Neigh 0{e'}\subseteq\cV$. This defines a
topology for which one checks easily that addition and scalar multiplication
are continuous ($\Field$ being equipped with the discrete topology). Actually
$\Fmod\Field X$ is a linearly topologized vector space in the sense
of~\cite{Lefschetz42}: the topology is generated by neighborhoods of $0$ which
are linear subspaces (for instance, the $\Neigh 0{e'}$ we introduced
above). This topology is Hausdorff: for any $a\in\Fmod\Field X$, if $a\not=0$
one cant find a (linear) neighborhood of $0$ which does not contain $a$. In
particular, the specialization ordering is discrete (this is not a topology
``\`a la Scott'').

\paragraph{Convergence and completeness.}
A net of $\Fmod\Field X$ if a family $(a(\gamma))_{\gamma\in\Gamma}$ of
elements $\Fmod\Field X$ indexed by a directed set $\Gamma$. Such a net
converges to $a\in\Fmod\Field X$ if, for any open linear subspace $\cV$ of
$\Fmod\Field X$ there is $\gamma\in\Gamma$ such that $\forall\delta\in\Gamma\
\delta\geq\gamma\Implies a(\delta)-a\in\cV$. If this holds, $a$ is unique
($\Fmod\Field X$ is Hausdorff). A net $(a(\gamma))_{\gamma\in\Gamma}$ is Cauchy
if for any open linear subspace $\cV$ of $\Fmod\Field X$, there exists
$\gamma\in\Gamma$ such that $\forall\delta\in\Gamma\ \delta\geq\gamma\Implies
a(\delta)-a(\gamma)\in\cV$. Using crucially the fact that $\Fin X=\Biorth{\Fin
  X}$, one can prove that any Cauchy net converges ($\Fmod\Field X$ is
complete).

\section{The basic finiteness structure}
We set
\begin{eqnarray*}
  \Snfinite &=& \Orth{\{\Up s\St s\in\Sterms\}}\\
  &=&\{e\subseteq\Sterms\St\forall s\in\Sterms\quad e\cap\Up s\text{ is
  finite}\}\,. 
\end{eqnarray*}
One defines similarly $\Pnfinite\subseteq\Part{\Pterms}$ as
$\Pnfinite=\{E\subseteq\Pterms\St\forall S\in\Pterms\quad E\cap\Up S\text{ is
  finite}\}$. This defines finiteness structures on $\Sterms$ and $\Pterms$. We
consider therefore $(\Sterms,\Snfinite)$ as a finiteness space that we simply
denote as $\Snfinite$.  To get a better grasp of the topology of the vector
space $\Fmod\Field\Snfinite$, we must make a first observation. We express
everything for $\Sterms$ for notational convenience, but obviously what we do
can be transposed to $\Pterms$ without any difficulty.

\begin{lemma}
  A subset $e'$ of $\Sterms$ belongs to $\Orth\Snfinite$ iff there are finitely
  many elements $s_1,\dots,s_n\in\Sterms$ such that 
  \begin{equation*}
    e'\subseteq\Up{s_1}\cup\dots\cup\Up{s_n}=\Up{\{s_1,\dots,s_n\}}\,.
  \end{equation*}
\end{lemma}
\Beginproof
The ``if'' part is trivial, let us check the ``only if'' part. The only
property of the order relation on simple terms that we need is the fact that
each set $\Down s$ is finite (Lemma~\ref{lemma:down-finite}).

Assume that there exists $e'\in\Orth\Snfinite$ such that
$e'\subseteq\Up{\{s_1,\dots,s_n\}}$ never holds. The set $e'$ cannot be empty,
so let $u_1\in e'$. Since $\Down{u_1}$ is finite, we cannot have
$e'\subseteq\Up{\Down{u_1}}$. So let $u_2\in e'\setminus\Up{\Down{u_1}}$.
Again, $\Down{u_2}$ being finite, we cannot have
$e'\subseteq\Up{\Down{u_1}}\cup\Up{\Down{u_2}}$. In that way, we construct an
infinite sequence $u_1,u_2\dots$ of elements of $e'$ such that for each $i$,
$u_{i+1}\in e'\setminus(\Up{\Down{u_1}}\cup\dots\cup\Up{\Down{u_i}})$; in
particular, the $u_i$'s are pairwise distinct, but we can say better: let $i<j$
and assume that $\Down{u_i}\cap\Down{u_j}\not=\emptyset$. Then
$u_j\in\Up{\Down{u_i}}$ and this is impossible. Let us set
$e=\{u_1,u_2,\dots\}$. For any $s\in\Sterms$, it follows from the disjointness
of the sets $\Down{u_i}$ that $e\cap\Up s$ has at most one element and is
therefore finite, so that $e\in\Snfinite$. But $e$ has an infinite intersection
with $e'$ (namely $e$), and this contradicts our hypothesis that
$e'\in\Orth\Snfinite$.
\Endproof

Therefore the topology of $\Fmod\Field\Snfinite$ is generated by the basic
neighborhoods $\Snneigh{s_1,\dots,s_n}=\{u\in\Fmod\Field\Snfinite\St\Supp
u\cap\Up{s_1}=\dots=\Supp u\cap\Up{s_n}=\emptyset\}$, where $s_1,\dots,s_n$ is
an arbitrary finite family of elements of $\Sterms$. Observe that these $s_i$'s
can be assumed to be minimal in $\Sterms$. An element $s$ of $\Sterms$ is
minimal for the order relation we have defined iff $s$ is normal, or reduces
only to $0$. A typical non-normal minimal term is $\Linapp{\Abs xy}{z}$, where
$y$ and $z$ are distinct variables.

The main purpose of these definitions is to give meaning to a normalization
function on vectors. Consider indeed an arbitrary linear combinations of
resource lambda-terms, $a=\sum_{a\in\Sterms}a_ss\in\Imod\Field\Sterms$. We
would like to set $\NF(a)=\sum_{s\in\Sterms}a_s\NF(s)$. But there could
perfectly exist normal elements $s_0\in\Sterms$ such that, for infinitely many
$s\in\Sterms$, $s_0\in\NF(s)$ and $a_s\not=0$. If this is the case, we cannot
normalize $a$ because infinite sums are not allowed in $\Field$ which is an
arbitrary field\footnote{Of course, one could also consider infinite sums if
  the coefficients were real or complex numbers but this will be the object of
  further studies.}. As a typical example of this situation, consider
$a=x+\Linapp{\Abs xx}{x}++\Linapp{\Abs xx}{(\Linapp{\Abs xx}{x})}+\cdots$ All
the terms of this sum reduce to the same term $x$ and hence $\NF(a)$ is not
defined.

\begin{proposition}\label{prop:snfinite-normalize}
  The map $\NF$ given by $\NF(a)=\sum_{s\in\Sterms}a_s\NF(s)$ is well defined,
  linear and continuous from the topological vector space
  $\Fmod\Field\Snfinite$ to itself.
\end{proposition}
\Beginproof Given $s\in\Sterms$, we have $\Supp{\NF(s)}\subseteq\Down s$. So,
since $\Supp a\in\Snfinite$, for any $s_0\in\Nsterms$, there are only finitely
many $s\in\Supp a$ such that $s_0\in\Supp{\NF(s)}$. So the sum above makes
sense, it can be written
\begin{equation*}
  \NF(a)=\sum_{s_0\in\Nsterms}
    \Bigl(\sum_{\Biind{s\in\Supp a}{s_0\in\Down s}}a_s\NF(s)_{s_0}\Bigr)s_0\,.
\end{equation*}
All the elements of $\Supp{\NF(a)}$ being minimal, this set obviously belongs
to $\Snfinite$.

The map $\NF$ defined in that way is obviously linear, we must just check that
it is continuous at $0$ but this is easy; indeed, if
$V=\Snneigh{s_1,\dots,s_n}$ is a basic neighborhood of $0$ then, by definition
of $\Snneigh{s_1,\dots,s_n}$, if $t\in\Sterms$ satisfies $t\in V$, this means
that $t\notin\Up{s_i}$ for each $i$, and hence for no $i$ we can have
$s_i\in\NF(t)$. Therefore $\NF(t)\in V$.
\Endproof

We can also extend the $\Redonez$ reduction relation to $\Fmod\Field\Snfinite$
in a completely ``free\footnote{In the sense that each summand can be reduced
  independently from the others.}''  way. Indeed let
$a\in\Fmod\Field\Snfinite$. If one writes $a=\sum_{i\in\Nat}\alpha_is_i$ with
$s_i\in\Sterms$ and with the sole restriction (for this sum to make sense at
all) that for each $s\in\Sterms$ there are only finitely $i$'s such that
$s_i=s$ and if, for each $i\in\Nat$, one chooses arbitrarily
$a(i)\in\Fmod\Nat\Sterms$ such that $s_i\Rel{\Redonez} a(i)$, then the sum
$b=\sum_{i\in\Nat}\alpha_ia(i)$ always makes sense, and belongs to
$\Fmod\Field\Snfinite$ (these facts result from the very definition of
$\Snfinite$). In that case we write $a\Rel\Redonez b$, and we denote by $\Red$
the transitive closure of $\Redonez$.
\begin{proposition}
  The relation $\Red$ is confluent on $\Fmod\Field\Snfinite$.
\end{proposition}
\Beginproof 
(Sketch) Use the confluence of $\Redone$ on $\Fmod\Nat\Sterms$ and the
following observation: given two finite families $(\alpha_i)_{i\in I}$ and
$(\beta_j)_{j\in J}$ of elements of $\Field$ such that
$\sum\alpha_i=\sum\beta_j $, one can find a family $(\gamma_{i,j})_{i\in I,j\in
  J}$ of elements of $\Field$ such that $\forall i\
\alpha_i=\sum_j\gamma_{i,j}$ and $\forall j\ \beta_j=\sum_i\gamma_{i,j}$.
\Endproof
One has to be aware that this ``reduction'' relation has strange properties and
can hardly be expected to normalize in a standard sense. For instance if
$s\Rel\Redone a_1$ and $s\Rel\Redone a_2$ where $a_1,a_2\in\Fmod\Nat\Sterms$
are distinct, then $0=s-s\Rel\Red a_1-a_2\not=0$ and the reduction can go on
after that. See~\cite{Vaux07,Vaux08} for more explanations. It makes sense
nevertheless to define the associated equivalence relation (the symmetric
closure of $\Red$) that we denote as $\Redeq$.

\begin{proposition}
  Let $a,b\in\Fmod\Field\Sterms$ be such that $a\Redeq b$. Then $\NF(a)=\NF(b)$.
\end{proposition}
\Beginproof
It suffices to show that $a\Rel\Redonez b\Implies\NF(a)=\NF(b)$ and this is
easy because $s\Rel\Redone c\Implies\NF(s)=\NF(c)$.
\Endproof
The converse implication does not hold because reducing an element
$a\in\Fmod\Field\Sterms$ to $\NF(a)$ can require an infinite number of
$\Redonez$ steps. But one can always exhibit sequences $a=a(1)\Rel\Redonez
a(2)\Rel\Redonez a(3)\cdots$ with $\lim_{n\to\infty}a(n)=\NF(a)$ (in the sense
of the topology of $\Fmod\Field\Snfinite$).

\begin{remark}
  It is not difficult to see that, given a finiteness space $X$, the
  topological space $\Fmod\Field X$ is metrizable (ie.~its topology can be
  defined by a distance) iff there exists an increasing sequence
  $(e'(n))_{n\in\Nat}$ of elements of $\Orth{\Fin X}$ such that $\forall
  e'\in\Orth{\Fin X}\,\exists n\in\Nat\ \ e'\subseteq e'(n)$. It is also
  interesting to observe that, when interpreting linear logic in finiteness
  spaces (see~\cite{Ehrhard00b}), one builds quite easily spaces which have not
  this property: for instance the interpretation of $!?1$ (the formula $1$
  being interpreted by the finiteness space $(\{*\},\{\emptyset,\{*\}\})$) is
  not metrizable.

  So the space $\Fmod\Field\Snfinite$ is metrizable: choose an enumeration
  $s_1,s_2,\dots$ of $\Sterms$ and, given $a,a'\in\Fmod\Field\Snfinite$, define
  $d(a,a')=0$ if $a=a'$, and $d(a,a')=2^{-n}$ where $n$ is the least integer
  such that $\Up{s_n}\cap\Supp{a-a'}\not=\emptyset$. This distance generates
  the topology we have defined, but presenting this space as a metric space
  would be unnatural, because there is (apparently) no canonical choice of such
  a distance (it depends on a completely arbitrary enumeration of $\Sterms$).

  A last interesting observation is that the subspace of $\Fmod\Field\Snfinite$
  spanned by the normal resource term is linearly compact\footnote{This notion
    is defined in~\cite{Lefschetz42}; it is a notion of compactness adapted to
    this setting.}, so that $\NF$ can be seen as a projection onto a linearly
    compact subspace.
\end{remark}

\subsection{Dealing with free variables}\label{sec:-finiteness-free-variables}
The finiteness space $\Snfinite$ allows to give meaning to normalization as
shown by Proposition~\ref{prop:snfinite-normalize}, but we would also like to
deal with elements of $\Snfinite$ (or of $\Fmod\Field\Snfinite$) as if they
were lambda-terms. However, nothing prevents an element $e$ of $\Snfinite$ of
containing infinitely many free variables. The set $\FV(e)$ can even be the set
of all variables: take for $e$ the set of all variables itself! It would be
hard to define $\beta$-reduction if we have to deal with such objects.

Fortunately the solution to this problem is quite easy. Let
$\cS\subseteq\Sterms$ be the set of all subsets $e'$ of $\Sterms$ such that,
for each finite set $\xi$ of variables, there are only finitely many elements
$s$ of $e'$ such that $\FV(s)\subseteq\xi$.

\begin{lemma}
  $\Orth\cS=\{e\subseteq\Sterms\St\FV(e)\text{ is finite}\}$.
\end{lemma}
\Beginproof
The inclusion ``$\supseteq$'' is straightforward. So let $e\in\Orth\cS$.
Towards a contradiction, assume that $\FV(e)$ is infinite and let
$x_1,x_2\dots$ be a repetition-free enumeration of this set of variables. Let
$n_1=1$. Choose $s_1\in e$ such that $x_1\in\FV(s_1)$. Since $\FV(s_1)$ is
finite, we can find $n_2$ such that $\FV(s_1)\cap\{x_i\St i\geq
n_2\}=\emptyset$. Choose $s_2\in e$ such that $x_{n_2}\in\FV(s_{2})$, choose
$n_3$ such that $\FV(s_2)\cap\{x_i\St i\geq n_3\}=\emptyset$\dots{} In that way
we define a sequence $s_1,s_2,\dots$ of element of $e$ and a sequence
$y_1,y_2,\dots$ of variables such that $y_i\in\FV(s_j)$ iff $i=j$ (take
$y_i=x_{n_i}$). Then $e'=\{s_i\St i=1,2,\dots\}$ is an element of
$\cS$. Indeed, if $\xi$ is a finite set of variables, $\xi$ contains only a
finite number of $y_i$'s and hence there can be only finitely many $i$'s such
that $\FV(s_i)\subseteq\xi$. But $e\cap e'$ is infinite since $e'\subseteq e$,
whence the contradiction.
\Endproof
This is another instance of a general proof scheme used several times
in~\cite{Ehrhard00b} and generalized by Tasson and Vaux (see~\cite{Tasson09}).

We arrive to the final definition of our basic finiteness space: we set
$\Snfinitefv=\Snfinite\cap\Orth\cS=\Orth{(\{\Up s\St s\in\Sterms\}\cup\cS)}$
and therefore we have $\Biorth\Snfinitefv=\Snfinitefv$ so that $\Snfinitefv$ is
actually a finiteness space.

\section{Interpreting types}

With any type (of system F, see Section~\ref{sec:syntax-F}), we want to
associate a finiteness space whose web will be a subset of $\Sterms$. The
construction is based on the definition of \emph{saturated sets}
in~\cite{Krivine93}, so we shall call our finiteness spaces saturated as well.

Let $\Snfinitevar$ be the collection of all subsets of $\Sterms$ which are of
the shape $\Linapp{\Linapp{\Linapp{x}{\Prom{e_1}}}{\cdots}}{\Prom{e_n}}$ where
$x$ is a variable and $e_1,\dots,e_n\in\Snfinitefv$.

\subsection{Saturated finiteness space} 

A \emph{$\Sterms$-finiteness space} is a finiteness space $X$ such that $\Web
X\subseteq\Sterms$.  One says that such a space $X$ is \emph{saturated} if
$\Snfinitevar\subseteq\Fin X\subseteq\Snfinitefv$ and, whenever
$g,e,e_1,\dots,e_n\in\Snfinitefv$, one has (using the notations introduced
in~\ref{par:extended-syntax} and~\ref{par:diff-subst}) the implication
\begin{eqnarray}\label{eq:saturation-expansion}
  &&\Linapp{\Linapp{\Linapp{\Msubst g{\Prom e}x}
      {\Prom{e_1}}}{\cdots}}{\Prom{e_n}}\in\Fin X\nonumber\\
  &&\hspace{3em}\Implies
  \Linapp{\Linapp{\Linapp{\Linapp{\Abs xg}{\Prom e}}{\Prom{e_1}}}
    {\cdots}}{\Prom{e_n}}\in\Fin X\,.
\end{eqnarray}
Then one simply says that $X$ is a \emph{saturated finiteness space}. 


Given two $\Sterms$-finiteness spaces $X$ and $Y$, we construct a new one,
denoted as $\Impl XY$.

The web $\Web{\Impl XY}$ is the collection of all $t\in\Sterms$ such that
\begin{equation*}
  \forall e\in\Fin X\quad \Linapp t{\Prom e}\in\Fin Y\,.
\end{equation*}

Then we define $\Fin{\Impl XY}$ as the collection of all $g\subseteq\Web{\Impl
  XY}$ such that
\begin{equation*}
  \forall e\in\Fin X\quad \Linapp g{\Prom e}\in\Fin Y\,,
\end{equation*}
that is
\begin{equation*}
  \forall e\in\Fin X,\,\forall f'\in\Orth{\Fin Y}
  \quad \Linapp g{\Prom e}\cap f'\ \text{is finite.}
\end{equation*}

Given $e\in\Fin X$ and $f'\in\Orth{\Fin Y}$, let $\Preimpl
e{f'}=\{t\in\Sterms\St\Linapp t{\Prom e}\cap f'\not=\emptyset\}$.
\begin{proposition}
  If $X$ and $Y$ are $\Sterms$-finiteness spaces, then
  \begin{equation}\label{eq:finiteness-impl-orthogonal}
    \Fin{\Impl XY}=\Orth{\{\Preimpl e{f'}\St e\in\Fin X,\
      f'\in\Orth{\Fin Y}\}}\,
  \end{equation}
  so that $\Impl XY$ is a $\Sterms$-finiteness space. If moreover $Y$ is
  saturated, then $\Impl XY$ is saturated as well.
\end{proposition}

\Beginproof
Let us check equation~(\ref{eq:finiteness-impl-orthogonal}), so let
$g\subseteq\Web{\Impl XY}$.

Assume first that $g\in\Fin{\Impl XY}$. Let $e\in\Fin X$ and $f'\in\Orth{\Fin
  Y}$. We know that $\Linapp g{\Prom e}\cap f'$ is finite. Let $t\in
g\cap(\Preimpl e{f'})$. This means that there exists $S_t\in\Prom e$ such that
$\Linapp t{S_t}\in f'$, that is, $\Linapp t{S_t}\in\Linapp g{\Prom e}\cap
f'$. But this latter set is finite, and the map $t\mapsto\Linapp t{S_t}$ is
injective, so the set $g\cap(\Preimpl e{f'})$ is finite as well.

Assume that $g\in\Orth{\{\Preimpl e{f'}\St e\in\Fin X\ \text{and}\
  f'\in\Orth{\Fin Y}\}}$ and let us show that $g\in\Fin{\Impl XY}$. So let
$e\in\Fin X$ and $f'\in\Orth{\Fin Y}$, we must show that $\Linapp g{\Prom
  e}\cap f'$ is finite. By definition of $\Preimpl e{f'}$, we have
\begin{equation*}
  \Linapp g{\Prom e}\cap f'
    = \bigcup_{t\in g\cap(\Preimpl e{f'})}(\Linapp t{\Prom e}\cap f')
\end{equation*}
and we conclude since $g\cap(\Preimpl e{f'})$ is finite, and, for $t\in g$, the
set $\Linapp t{\Prom e}\cap f'$ is finite since $g\subseteq\Web{\Impl XY}$
(remember the definition above of that set).

So $\Impl XY=(\Web{\Impl XY},\Fin{\Impl XY})$ is a finiteness space. Assume
that $Y$ is saturated and let us show that $\Impl XY$ is.

We have $\Snfinitevar\subseteq\Fin{\Impl XY}$: this results immediately from
$\Snfinitevar\subseteq\Fin Y$ and $\Fin X\subseteq\Snfinitefv$.

We have $\Fin{\Impl XY}\subseteq\Snfinitefv$: let $g\in\Fin{\Impl XY}$ and let
$t\in\Sterms$. We must show that $g\cap\Up t$ is finite, so assume towards a
contradiction that there are $t_1,t_2,\dots\in g$, pairwise distinct, and such
that $t_i\in\Up t$ for each $i$. This means that there are terms
$a_1,a_2,\dots\in\Fmod\Nat\Sterms$ such that $t_i\Rel\Red a_i$ and
$t\in a_i$ for each $i$. Let $x$ be an arbitrary variable, then
$\Linapp{t_i}x\Rel\Red\Linapp{a_i}x$ and $\Linapp tx\in\Supp{\Linapp{a_i}x}$
for each $i$, therefore $\Linapp gx\cap\Up{\Linapp tx}$ is infinite, which is
impossible because $\{x\}\in\Fin X$ (since $\Snfinitevar\subseteq\Web X$) and
$\Fin Y\subseteq\Snfinitefv$.

It remains to check that $\Fin{\Impl XY}$ satisfies
condition~(\ref{eq:saturation-expansion}), and this is straightforward.
\Endproof

\subsection{The ground space}

\begin{lemma}
  The finiteness space $(\Sterms,\Snfinitefv)$ is saturated.
\end{lemma}
\Beginproof
The only condition which is not obviously satisfied
is~(\ref{eq:saturation-expansion}). So let $g,e,e_1,\dots,e_n\in\Snfinitefv$
and assume that $\Linapp{\Linapp{\Linapp{\Msubst{g}{\Prom
        e}{x}}{\Prom{e_1}}}{\cdots}}{\Prom{e_n}}\in\Snfinitefv$. Let
$s\in\Sterms$, we must show that the intersection $\Up
s\cap\Linapp{\Linapp{\Linapp{\Linapp{\Abs xg}{\Prom
        e}}{\Prom{e_1}}}{\cdots}}{\Prom{e_n}}$ is finite. Let
$(s_i,S_i,S_{1,i},\dots,S_{n,i})_{i\in I}$ be a repetition free enumeration of
all the elements of $g\times\Prom
e\times\Prom{e_1}\times\cdots\times\Prom{e_n}$ such that
\begin{eqnarray*}
  &&t_i=\Linapp{\Linapp{\Linapp{\Linapp{\Abs x{s_i}}
        {{S_i}}}{{S_{1,i}}}}{\cdots}}{{S_{n,i}}}\\
  &&\hspace{5em}\in\Up
  s\cap\Linapp{\Linapp{\Linapp{\Linapp{\Abs xg}{\Prom
          e}}{\Prom{e_1}}}{\cdots}}{\Prom{e_n}}  
\end{eqnarray*}
Observe that all the free variables of the terms $t_i$ appear free in $s$ and
hence there are only finitely many such variables. So we can choose a variable
$y$ which is free in none of these terms. For each $i\in I$, we set
$v_i=\Linapp{\Linapp{\Linapp{y}{{S_{1,i}}}}{\cdots}}{{S_{n,i}}}\in\Sterms$, so
that $t_i=\Subst{v_i}{\Linapp{\Abs x{s_i}} {{S_i}}}{y}$. We can also assume
that $x$ occurs free or bound in none of the terms $S_i,S_{1,i},\dots,S_{n,i}$
(for all $i\in I$).  We apply Lemma~\ref{lemma:red-permutation}, considering
two cases.
\begin{itemize}
\item Either $x$ appears bound in $s$, and in that case we have
  $s=\Subst{v'}{\Linapp{\Abs x{s'}}{S'}}{y}$ for some $v',s'\in\Sterms$ and
  $S'\in\Pterms$ such that $v'\leq v_i$, $s'\leq s_i$ and $S'\leq S_i$ for each
  $i\in I$. We have $v'=\Linapp{\Linapp{\Linapp{y}{{S'_{1}}}}{\cdots}}{S'_{n}}$
  for $S'_1,\dots,S'_n\in\Pterms$ such that $S'_j\leq S_{j,i}$ for each
  $j\in\{1,\dots,n\}$ and $i\in I$. By the assumption that
  $g,e,e_1,\dots,e_n\in\Snfinitefv$ we see that the sets $\{s_i\St i\in I\}$,
  $\{S_i\St i\in I\}$, $\{S_{1,i}\St i\in I\}$,\dots,$\{S_{n,i}\St i\in I\}$
  are finite and so $\Up s\cap\Linapp{\Linapp{\Linapp{\Linapp{\Abs xg}{\Prom
          e}}{\Prom{e_1}}}{\cdots}}{\Prom{e_n}}$ is finite.
\item Or $x$ does not appear bound in $s$. Then for each $i\in I$ there exists
  $u_i\in\Sterms$ such that $u_i\in\Msubst{s_i}{S_i}x$ and
  $s\leq\Subst{v_i}{u_i}y$. In other words
  \begin{equation*}
    \forall i\in I\quad
    \Subst{v_i}{u_i}y\in\Up s\cap \Linapp{\Linapp{\Linapp{\Msubst{g}{\Prom
        e}{x}}{\Prom{e_1}}}{\cdots}}{\Prom{e_n}}
  \end{equation*}
  and hence by our assumption that $\Linapp{\Linapp{\Linapp{\Msubst{g}{\Prom
          e}{x}}{\Prom{e_1}}}{\cdots}}{\Prom{e_n}}\in\Snfinitefv$, the set
  $\{\Subst{v_i}{u_i}y\St i\in I\}$ is finite. Coming back to the definition of
  $v_i$, this means that the sets $\{u_i\St i\in I\}$, $\{S_{1,i}\St i\in
  I\}$,\dots,$\{S_{n,i}\St i\in I\}$ are finite. But for each $i\in I$, we know
  that there are only finitely many pairs $(w,W)\in\Sterms\times\Pterms$ such
  that $u_i\in\Msubst wWx$ by Lemma~\ref{lemma:msubst-reverse} and hence, since
  $u_i\in\Msubst{s_i}{S_i}x$, the sets $\{s_i\St i\in I\}$ and $\{S_i\St i\in
  I\}$ must be finite as well since $\{s_i\St i\in I\}$ is finite.\Endproof
\end{itemize}

\subsection{Inclusions and intersections of saturated finiteness spaces}
\label{sec:inclusion-sat}
Let $X$ and $Y$ be saturated finiteness spaces. We write $X\subseteq Y$ when
$\Web X\subseteq\Web Y$ and $\Fin X\subseteq\Fin Y$. This defines an order
relation on saturated finiteness spaces.

\begin{lemma}
  Let $(X_i)_{i\in I}$ be a family of saturated finiteness spaces. Then
  $\bigcap_{i\in I}X_i=(\bigcap_{i\in I}\Web{X_i},\bigcap_{i\in
    I}(\Fin{X_i}\cap\Part{\Web X_i}))$ is a saturated finiteness space, and it
  is le glb of the family $(X_i)_{i\in I}$.
\end{lemma}
\Beginproof
Let $X=\bigcap_{i\in I}X_i$. Let $e\subseteq\Web X=\bigcap_{i\in
  I}\Web{X_i}$. We assume that $e\in\Biorth{\Fin X}$ and we prove that
$e\in\Fin X$. Let $i\in I$, we must show that
$e\in\Fin{X_i}=\Biorth{\Fin{X_i}}$. So let $e'\subseteq\Web{X_i}$ and let us
show that $e\cap e'$ is finite. Since $e\in\Biorth{\Fin X}$, it will be
sufficient to show that $e'\in\Orth{\Fin X}$. So let $f\subseteq\Web X$ be such
that $f\in\Fin X$. In particular we have $f\in\Fin{X_i}$ and hence $e'\cap f$
is finite as required. So $X$ is a $\Sterms$-finiteness space.

Since $\Snfinitevar\subseteq\Fin{X_i}\subseteq\Snfinitefv$ holds for all $i\in
I$, and since $I$ is non empty, it is clear that
$\Snfinitevar\subseteq\Fin{X}\subseteq\Snfinitefv$.

Let $g,e,e_1,\dots,e_n\in\Snfinitefv$ be such that
$\Linapp{\Linapp{\Linapp{\Msubst g{\Prom e}x}
    {\Prom{e_1}}}{\cdots}}{\Prom{e_n}}\in\Fin X$. Then for each $i$ we have
$\Linapp{\Linapp{\Linapp{\Msubst g{\Prom e}x}
    {\Prom{e_1}}}{\cdots}}{\Prom{e_n}}\in\Fin{X_i}$ and hence
$\Linapp{\Linapp{\Linapp{\Linapp{\Abs xg}{\Prom e}}{\Prom{e_1}}}
  {\cdots}}{\Prom{e_n}}\in\Fin{X_i}$ and therefore
$\Linapp{\Linapp{\Linapp{\Linapp{\Abs xg}{\Prom e}}{\Prom{e_1}}}
  {\cdots}}{\Prom{e_n}}\in\Fin{X}$.
\Endproof

\section{Taylor expansion in an algebraic system F}

\subsection{Syntax of the algebraic system F}\label{sec:syntax-F}
The types are defined as usual: one has type variables $\phi,\psi\dots$, and if
$A$ and $B$ are types, so are $\Impl AB$ and $\Tforall\phi A$. We adopt the
Curry style for presenting system F, so that our terms are ordinary
lambda-terms, with the additional possibility of linearly combining terms, with
coefficients in $\Field$. More precisely, we define the set $\Alambda$ of
lambda-terms with coefficients in $\Field$ as follows:
\begin{itemize}
\item if $x$ is a variable then $x\in\Alambda$;
\item if $M\in\Alambda$ and $x$ is a variable, then $\Abs xM\in\Alambda$;
\item if $M\in\Alambda$ and $Q\in\Fmodfin\Field\Alambda$ then $\App
  MQ\in\Alambda$.
\end{itemize}
For $Q,R\in\Fmodfin\Field\Alambda$, we set $\Abs xQ=\sum_{M\in\Alambda}Q_M\Abs
xM$ and $\App QR=\sum_{M\in\Alambda}Q_M\App MR$. Observe that these two sums
are finite because $Q$ is a finite linear combination of terms. In other word,
abstraction is linear and application is left-linear (but not right-linear). We
give now the typing rules for terms belonging to $\Alambda$. A typing context
$\Gamma$ is as usual a finite partial function from variables to types.
\begin{center}
  \AxiomC{}
  \UnaryInfC{${\Gamma,x:A}\Seq{x:A}$}
  \DisplayProof
  \quad
  \AxiomC{${\Gamma,x:A}\Seq{M:B}$}
  \UnaryInfC{$\Gamma\Seq{\Abs xM:\Impl AB}$}
  \DisplayProof
\end{center}
\begin{center}
  \AxiomC{$\Gamma\Seq{M:\Impl AB}$}
  \quad
  \AxiomC{$\Gamma\Seq{N_1:A}$\quad \dots\quad $\Gamma\Seq{N_n:A}$}
  \BinaryInfC{$\Gamma\Seq{\App M{(\alpha_1N_1+\cdots+\alpha_nN_n):B}}$}
  \DisplayProof
\end{center}
\begin{center}
  \AxiomC{$\Gamma\Seq{M:\Tforall\phi A}$}
  \UnaryInfC{$\Gamma\Seq{M:\Subst AB\phi}$}
  \DisplayProof
  \quad\quad
  \AxiomC{$\Gamma\Seq{M:A}$}
  \UnaryInfC{$\Gamma\Seq{M:\Tforall\phi A}$}
  \DisplayProof
\end{center}
with, for the last rule, the usual side condition that $\phi$ should not
occur free in the typing context $\Gamma$.

\subsection{Taylor expansion}
Given a term $M\in\Alambda$ (resp.~$Q\in\Fmod\Field\Alambda$), we define a
generally infinite linear combinations $\Tay M$ (resp.~$\Tay Q$) of elements of
$\Sterms$, with coefficients in $\Field$, as follows:
\begin{eqnarray*}
  \Tay x &=& x\\
  \Tay{(\Abs xM)} &=& \Abs x{(\Tay M)}\\
  \Tay{(\App MQ)} &=& \sum_{n\in\Nat}\frac 1
  {\Factor n}\Linapp{\Tay M}{(\Tay Q)^n}\\
  \Tay Q &=& \sum_{M\in\Alambda}Q_M\Tay Q
\end{eqnarray*}
where we use the conventions of~\ref{par:extended-syntax} for infinite linear
combinations of terms. Let us be more explicit. With any term $M\in\Alambda$,
we associate a linear combination $\Tay M$ of elements of $\Sterms$ which can
be written
\begin{equation*}
  \Tay M=\sum_{s\in\Sterms}\Tay M_ss
\end{equation*}
where $\Tay M_s\in\Field$ for each $s$, and similarly we define $\Tay
Q_s\in\Field$ for each $Q\in\Fmod\Field\Alambda$. Then these numbers are given
inductively by:
\begin{eqnarray*}
  \Tay x_s &=&
  \begin{cases}
    1 & \text{if $s=x$}\\
    0 & \text{otherwise}
  \end{cases}\\
  \quad\quad\quad
  \Tay{(\Abs xM)_s} &=&
  \begin{cases}
    \Tay M_t & \text{if $s=\Abs xt$}\\
    0 & \text{otherwise}
  \end{cases}\\
  \quad\quad\quad
  \Tay Q_s&=&\sum_{M\in\Alambda}Q_M\Tay M_s
\end{eqnarray*}
Last, $\Tay{(\App MQ)}_s=0$ if $s$ is not an application, and otherwise
\begin{eqnarray*}
  \Tay{(\App MQ)}_{\Linapp tT}
  &=&\left(\sum_{n\in\Nat}\frac 1
    {\Factor n}\Linapp{\Tay M}{(\Tay Q)^n}\right)_{\Linapp tT}\\
  &=&\sum_{n\in\Nat}\frac{\Tay M_t}{\Factor n}{(\Tay Q)^n_T}\\
  &=&\sum_{n\in\Nat}\frac{\Tay M_t}{\Factor n}
  {\left(\sum_{u\in\Sterms}\Tay Q_uu\right)^n_T}\\
  &=&\frac{\Tay M_t(\Tay Q)^T}{\Factor T}
  \end{eqnarray*}
where $\Factor T=\prod_{u\in\Sterms}\Factor{T(u)}$ and $(\Tay
Q)^T=\prod_{u\in\Sterms}(\Tay Q_u)^{T(u)}$ (see~\cite{EhrhardRegnier06a} for
more details on this kind of algebraic computations); remember that $T$ is a
finite multiset of elements of $\Sterms$ and that $T(u)\in\Nat$ is the
multiplicity of $u$ in $T$.

Given $M\in\Alambda$, we define a set $\Shape M\subseteq\Sterms$ as follows:
\begin{equation*}
  \Shape x=\{x\},\quad \Shape{\Abs xM}=\{\Abs xs\St s\in\Shape M\}
\end{equation*}
\begin{eqnarray*}
  &&\Shape{\App M{(\alpha_1N_1+\cdots+\alpha_nN_n)}}\\
  &&\quad=\{\Linapp s{(t_1\cdots t_p)}\St  s\in\Shape M\\
  &&\quad\quad\quad\text{and}\
  t_1,\dots,t_p\in\Shape{N_1}\cup\cdots\cup\Shape{N_n}\}\,. 
\end{eqnarray*}

The following property follows readily from these definitions.
\begin{lemma}\label{lemma:support-shape}
  Let $M\in\Alambda$ and $s\in\Sterms$. If $\Tay M_s\not=0$ then $s\in\Shape
  M$. 
\end{lemma}

\subsection{The standard case: coherence}
When the algebraic lambda-term $M$ is a standard lambda-term, that is an
element of $\Alambda$ where all the linear combinations
$\alpha_1N_1+\cdots+\alpha_pN_p$ are trivial in the sense that all $\alpha_i$'s
are equal to $0$ but one which is equal to $1$, we showed
in~\cite{EhrhardRegnier06a} that the Taylor expansion can be written $\Tay
M=\sum_{s\in\Shape M}\frac 1{m(s)}s$ where $m(s)\in\Nat\setminus\{0\}$ is an
integer which depends only on $s$ (in other words $\Tay M_s$ depends on $M$ in
a very simple way: $\Tay M_s=0$ if $s\notin\Shape M$, and otherwise $\Tay
M_s=1/m(s)$). Moreover the various elements of $\Shape M$ cannot overlap during
their reduction, in the sense that if $s,t\in\Shape M$ are distinct then
$\NF(s)\cap\NF(t)=\emptyset$. This is proven by introducing a binary symmetric
but not reflexive coherence relation, observing that each set $\Shape M$ is a
clique for this coherence relation and proving that $\NF$ can be seen as a
stable and linear function on this coherence space (in the sense
of~\cite{Girard86}).

These properties are lost in the present setting and superpositions can occur
and even lead to infinite sums, as in the Taylor expansion (that we do not
compute here) of the term $M=\App\Theta{\Abs x{(x+z)}}$ where $z$ is a variable
$\not=x$ and $\Theta$ is the Turing fixpoint combinator (reducing $M$ leads to
terms of the shape $nz+M$ for all $n\in\Nat$). This superposition of elementary
normal forms is controlled by the finiteness structures, but this is possible
only in a typed setting (here, second order types).

\subsection{Finiteness of the Taylor expansions in system F}
\subsubsection{Interpreting types}
A type valuation is a map $\cI$ which associates a saturated finiteness space
$\cI(\phi)$ with any type variable $\phi$. By induction on type $A$ we define,
for all valuation $\cI$, a saturated finiteness space $\Tsem A\cI$ in a fairly
standard way: $\Tsem\phi\cI=\cI(\phi)$, $\Tsem{(\Impl AB)}\cI=\Impl{\Tsem
  A\cI}{\Tsem B\cI}$ and $\Tsem{(\Tforall\phi A)}\cI=\bigcap_{X\in\Satfs}\Tsem
A{\cI[\phi\mapsto X]}$ where $\Satfs$ is the class of all saturated finiteness
spaces (remember that the intersection of saturated finiteness spaces is
defined in Section~\ref{sec:inclusion-sat}).

\subsubsection{The fundamental property}
Our goal is to prove that, if $\Gamma\Seq{M:A}$, then $\Shape M\in\Fin{\Tsem
  A\cI}$ for any valuation $\cI$. Of course this property cannot be proven in
that form and a more general statement is needed.

\begin{proposition}\label{prop:interpretation}
  Let $\Gamma=(x_1:A_1,\dots,x_n:A_n)$ be a typing context.  Assume that
  $\Gamma\Seq M:B$, where $M\in\Alambda$ and $B$ and the $A_i$'s are second
  order types. Let $\cI$ be a valuation. Let
  $e_1\in\Fin{\Tsem{A_1}\cI}$,\dots,$e_n\in\Fin{\Tsem{A_n}\cI}$ be sets of
  simple terms and let $f=\Shape M$. Then
  $\Msubst{f}{\Prom{e_1},\dots,\Prom{e_n}}{\List x1n}\in\Fin{\Tsem B\cI}$.
\end{proposition}
\Beginproof
Adaptation from the proof of strong normalization of system~F
in~\cite{Krivine93}, see the Appendix.
\Endproof

By Lemma~\ref{lemma:support-shape}, this shows in particular that, if
$M\in\Alambda$ is typeable in system F, then $\Tay M\in\Fmod\Field\Snfinitefv$
so that we can reduce the infinitely many resource terms appearing in this
expansion without creating any infinite superimposition of terms, whatever be
the choices we make in this process. Of course, one can also prove that
$\NF(\Tay M)=\Tay{M_0}$ where $M_0$ is the normal form of $M$, but this is not
straightforward.



\section*{Conclusion}
Following the line of ideas initiated
in~\cite{EhrhardRegnier02,EhrhardRegnier06a,EhrhardRegnier06b}, we considered
the resource lambda-calculus as an algebraic setting where various (algebraic,
differential\dots) extensions of the lambda-calculus can be interpreted. In
this setting, the elementary points of the interpretation (the simple resource
terms) are considered as base vectors and, in sharp contrast with denotational
semantics, have their own completely finite dynamics. We introduced topologies
for controlling their global behavior during reduction and avoiding the
appearance of infinite coefficients: linear combinations of resource terms are
organized as Hausdorff and complete topological vector spaces associated with
types. By a rather standard reducibility argument, we proved that the Taylor
expansion of any term of an algebraic extension of system F belongs to the
vector space interpretation of its type, but of course these vector spaces
contain many elements which are not Taylor expansions of such terms. 

For instance, given $a\in\Fmod\Field{\Impl XY}$, it is not difficult to define
$a'\in\Fmod\Field{\Impl X{(\Impl XY)}}$, the derivative of $a$ (which is linear
in its first parameter of type $X$). Saying that $a'$ is linear means that
$\Linapp{a'}{x^n}\Rel\Redeq 0$ for $n\not=1$, where $x$ is an arbitrary
variable. One can show that this operation can be reversed (under a necessary
and sufficient condition), so that it makes sense to compute ``primitives'' of
resource terms and it is certainly a fascinating challenge to understand the
operational meaning of this operation.

\bibliographystyle{alpha} \bibliography{newbiblio}

\newcommand{\SortNoop}[1]{}
\begin{thebibliography}{BCL99}

\bibitem[AD08]{ArrighiDowek08}
Pablo Arrighi and Gilles Dowek.
\newblock Linear-algebraic lambda-calculus: higher-order, encodings, and
  confluence.
\newblock In Andrei Voronkov, editor, {\em RTA}, volume 5117 of {\em Lecture
  Notes in Computer Science}, pages 17--31. Springer, 2008.

\bibitem[BCL99]{BoudolCurienLavatelli99}
G\'erard Boudol, Pierre-Louis Curien, and Carolina Lavatelli.
\newblock A semantics for lambda calculi with resource.
\newblock {\em Mathematical Structures in Computer Science}, 9(4):437--482,
  1999.

\bibitem[Bou93]{Boudol93}
G\'erard Boudol.
\newblock The lambda calculus with multiplicities.
\newblock Technical Report 2025, INRIA Sophia-Antipolis, 1993.

\bibitem[Ehr05]{Ehrhard00b}
Thomas Ehrhard.
\newblock Finiteness spaces.
\newblock {\em Mathematical Structures in Computer Science}, 15(4):615--646,
  2005.

\bibitem[EL09]{EhrhardLaurent08}
Thomas Ehrhard and Olivier Laurent.
\newblock Interpreting a finitary pi-calculus in differential interaction nets.
\newblock {\em Information and Computation}, 2009.
\newblock To appear.

\bibitem[ER03]{EhrhardRegnier02}
Thomas Ehrhard and Laurent Regnier.
\newblock The differential lambda-calculus.
\newblock {\em Theoretical Computer Science}, 309(1-3):1--41, 2003.

\bibitem[ER06a]{EhrhardRegnier06b}
Thomas Ehrhard and Laurent Regnier.
\newblock {B}\"ohm trees, {K}rivine machine and the {T}aylor expansion of
  ordinary lambda-terms.
\newblock In Arnold Beckmann, Ulrich Berger, Benedikt L\"{o}we, and John~V.
  Tucker, editors, {\em Logical Approaches to Computational Barriers}, volume
  3988 of {\em Lecture Notes in Computer Science}, pages 186--197.
  Springer-Verlag, 2006.

\bibitem[ER06b]{EhrhardRegnier06d}
Thomas Ehrhard and Laurent Regnier.
\newblock Differential interaction nets.
\newblock {\em Theoretical Computer Science}, 364(2):166--195, 2006.

\bibitem[ER08]{EhrhardRegnier06a}
Thomas Ehrhard and Laurent Regnier.
\newblock Uniformity and the {T}aylor expansion of ordinary lambda-terms.
\newblock {\em Theoretical Computer Science}, 403(2-3):347--372, 2008.

\bibitem[Gir86]{Girard86}
Jean-Yves Girard.
\newblock The system {F} of variable types, fifteen years later.
\newblock {\em Theoretical Computer Science}, 45:159--192, 1986.

\bibitem[Kfo00]{Kfoury00}
Assaf~J. Kfoury.
\newblock A linearization of the lambda-calculus.
\newblock {\em Journal of Logic and Computation}, 10(3):411--436, 2000.

\bibitem[Kri93]{Krivine93}
Jean-Louis Krivine.
\newblock {\em Lambda-Calculus, Types and Models}.
\newblock Ellis Horwood Series in Computers and Their Applications. Ellis
  Horwood, 1993.
\newblock Translation by Ren{\'e} Cori from French 1990 edition (Masson).

\bibitem[Lef42]{Lefschetz42}
Solomon Lefschetz.
\newblock {\em Algebraic topology}.
\newblock Number~27 in American mathematical society colloquium publications.
  American Mathematical Society, 1942.

\bibitem[PT09a]{PaganiTasson09}
Michele Pagani and Christine Tasson.
\newblock {The Taylor Expansion Inverse problem in Linear Logic}.
\newblock In {\em Proceedings of the 24th Annual IEEE Symposium on Logic in
  Computer Science}, pages 222--232. IEEE Computer Society, 2009.

\bibitem[PT09b]{PaganiTranquilli09}
Michele Pagani and Paolo Tranquilli.
\newblock {Parallel Reduction in Resource Lambda-Calculus}.
\newblock In Zhenjiang Hu, editor, {\em APLAS}, volume 5904 of {\em Lecture
  Notes in Computer Science}, pages 226--242. Springer, 2009.

\bibitem[Tas09]{Tasson09}
Christine Tasson.
\newblock {\em S\'emantiques et syntaxes vectorielles de la logique
  lin\'eaire}.
\newblock Th\`ese de doctorat, Universit\'e Paris Diderot -- Paris 7, 2009.

\bibitem[Tra08]{Tranquilli08}
Paolo Tranquilli.
\newblock {Intuitionistic Differential Nets and Lambda-Calculus}.
\newblock {\em Theoretical Computer Science}, 2008.
\newblock To appear.

\bibitem[Vau05]{Vaux05}
Lionel Vaux.
\newblock The differential lambda-mu calculus.
\newblock {\em Theoretical Computer Science}, 379(1-2):166--209, 2005.

\bibitem[Vau07]{Vaux07}
Lionel Vaux.
\newblock On linear combinations of lambda-terms.
\newblock In {\em Term Rewriting and Applications}, volume 4533 of {\em Lecture
  Notes in Computer Science}, pages 374--388. Springer-Verlag, 2007.

\bibitem[Vau08]{Vaux08}
Lionel Vaux.
\newblock The algebraic lambda-calculus.
\newblock {\em Mathematical Structures in Computer Science}, 2008.

\end{thebibliography}

\section*{Appendix: proof of proposition~\ref{prop:interpretation}}
We adopt the following notational convention: if $g\in\Fin{\Tsem C\cI}$ for
some type $C$ then we use $g'$ to denote the set
$\Msubst{g}{\Prom{e_1},\dots,\Prom{e_n}}{\List x1n}$.

The proof is by induction on the typing derivation of
$x_1:A_1,\dots,x_n:A_n\Seq M:B$ (the statement that we prove by induction is
universally quantified in $\cI$ and in the $e_i$'s).

Assume first that $M=x_i$ and that the derivation consists of the axiom
\begin{center}
  \AxiomC{}
  \UnaryInfC{$\Gamma\Seq x_i:A_i$}
  \DisplayProof
\end{center}
We have $f=\{x_i\}$ and hence
\begin{eqnarray*}
  &&\Msubst{f}{\Prom{e_1},\dots,\Prom{e_n}}{\List x1n}\\
  &&\quad=
  \cup\{\Msubst{x_i}{\List S1n}{\List x1n}\St \forall j\ S_j\in\Prom{e_j}\}\\
  &&\quad=e_i\in\Fin{\Tsem{A_i}\cI}\,.
\end{eqnarray*}

Assume that $M=\App N{(\alpha_1L_1+\cdots+\alpha_pL_p)}$ where $N,\List
L1p\in\Alambda$ and that the derivation ends with
\begin{center}
  \AxiomC{$\Gamma\Seq{N:\Impl AB}$}
  \quad
  \AxiomC{$\Gamma\Seq{L_1:A}$\ \dots\ $\Gamma\Seq{L_p:A}$}
  \BinaryInfC{$\Gamma
    \Seq{\App N{(\alpha_1L_1+\cdots+\alpha_pL_p):B}}$}
  \DisplayProof
\end{center}
We set $Q=\alpha_1L_1+\cdots+\alpha_pL_p\in\Fmodfin\Field\Alambda$ and
$h=\Shape Q$.

Let $g=\Shape N$ and let $h_i=\Shape{L_i}$ for $i=1,\dots,p$. By inductive
hypothesis, we have $g'\in\Fin{\Tsem{\Impl AB}\cI}$ and
$h_j'\in\Fin{\Tsem{A}\cI}$ for $j=1,\dots,p$. Since $h\subseteq
h_1\cup\cdots\cup h_p$ and hence $h'\subseteq h'_1\cup\cdots\cup
h'_p\in\Fin{\Tsem A\cI}$ (remember from Section~\ref{sec:finiteness-spaces}
that $\Tsem A\cI$ is closed under finite unions).

By definition of $\Tsem{\Impl AB}\cI$, we have therefore
$\Linapp{g'}{\Prom{(h')}}\in\Fin{\Tsem B\cI}$. Since $f=\Shape{\App NQ}=\Linapp
g{\Prom h}$, we have $f'=\Linapp{g'}{\Prom{(h')}}$ and we conclude for that
case.

Assume that $M=\Abs xN$ where $N\in\Alambda$ and that the derivation ends with
\begin{center}
  \AxiomC{$\Gamma,x:B\Seq N:C$}
  \UnaryInfC{$\Gamma\Seq{\Abs xN:\Impl BC}$}
  \DisplayProof
\end{center}
so that $A=(\Impl BC)$. Let $g=\Shape N$, we have $f=\Abs xg$ and hence
$f'=\Abs x{g'}$ (as usual we assume that $x$ is different from all the $x_i$'s
and does not occur free in the $e_i$'s; this is possible because
$e_i\in\Snfinitefv$ and hence $\FV(e_i)$ is finite for each $i$, see
Section~\ref{sec:-finiteness-free-variables}) and we must prove that $\Abs
x{g'}\in\Fin{\Tsem{\Impl BC}\cI}$. Let $e\in\Fin{\Tsem B\cI}$, we must prove
that $\Linapp{\Abs x{g'}}{\Prom e}\in\Fin{\Tsem C\cI}$. Since $\Tsem C\cI$ is a
saturated finiteness space, it suffices to prove that $\Msubst{g'}{\Prom
  e}x=\Msubst{g}{\Prom{e_1},\dots,\Prom{e_n},\Prom e}{\List x1n,x}\in\Fin{\Tsem
  C\cI}$ and this results from the inductive hypothesis.

Assume that the derivation ends with
\begin{center}
  \AxiomC{$\Gamma\Seq{M:\Tforall\phi A}$}
  \UnaryInfC{$\Gamma\Seq{M:\Subst AB\phi}$}
  \DisplayProof
\end{center}
By inductive hypothesis we have
\begin{equation*}
f'\in\Fin{\Tsem{\Tforall\phi A}\cI}
=\bigcap_{X\in\Satfs}\Tsem A{\cI[\phi\mapsto X]}
\subseteq\Fin{\Tsem A{\cI[\phi\mapsto\Tsem B\cI]}}
\end{equation*}
and we conclude because this finiteness space is $\Tsem{\Subst AB\phi}\cI$
(straightforward proof by induction on types).

Last assume that the proof ends with
\begin{center}
  \AxiomC{$\Gamma\Seq{M:A}$}
  \UnaryInfC{$\Gamma\Seq{M:\Tforall\phi A}$}
  \DisplayProof
\end{center}
and remember that $\phi$ cannot occur free in $\Gamma$. Given a saturated
finiteness space $X$ we set $\cI_X=\cI[\phi\mapsto X]$. Our assumption on the
$e_i$'s is that $e_i\in\Fin{\Tsem{B_i}\cI}$ for each $i$. Let $X$ be a saturated
finiteness space.  Since $\phi$ does not occur free in $\Gamma$, we have
$e_i\in\Fin{\Tsem{B_i}{\cI_X}}$ and hence by the inductive hypothesis we have
$f'\in\Fin{\Tsem{A}{\cI_X}}$. Since this holds for each $X$, we have
$f'\in\Fin{\Tsem{\Tforall\phi A}{\cI}}$.

\end{document}